\documentclass[twocolumn,showpacs,preprintnumbers,amsmath,amssymb,showkeys]{revtex4}

\usepackage{graphicx}
\usepackage{dcolumn}
\usepackage{bm}

\begin{document}

\preprint{\textbf{}}

\title{The counterion condensation of differently flexible polyelectrolyte aqueous
solutions in the dilute and semidilute regime}

\author{D. Truzzolillo}
\author{F. Bordi}
\author{C. Cametti}%
 \email{cesare.cametti@roma1.infn.it}
\author{S. Sennato}
\affiliation{%
Dipartimento di Fisica, Universita' di Roma "La Sapienza"
, Piazzale A. Moro 5, I-00185 Rome (Italy)\\
and CNR INFM-SOFT, Unita' di Roma1.
}%


\date{\today}

\begin{abstract}
The low-frequency limit of the electrical conductivity (d.c.
conductivity) of differently flexible polyions in aqueous
solutions has been measured over an extended polyion
concentration range, covering both the dilute and semidilute
(entangled and unentangled) regime, up to the concentrated
regime. The data have been analyzed taking into account the
different flexibility of the polymer chains according to the
scaling theory of polyion solutions, in the case of flexible
polyions, and according to the Manning model, in the case of
rigid polyions. In both cases, the fraction $f$ of free
counterions, released into the aqueous phase from the ionizable
polyion groups, has been evaluated and its dependence on the
polyion concentration determined. Our results show that the
counterion condensation follows at least three different regimes
in dependence on the polyion concentration. The fraction $f$ of
free counterions remains constant only in the semidilute regime
(a region that we have named \textit{the Manning regime}), while
there is a marked dependence on the polyion concentration both
in the dilute and in the concentrated regime. These results are
briefly discussed at the light of the aqueous ionic solutions.
\end{abstract}

\maketitle

\section{Introduction}
Polyelectrolytes are macromolecules that, in appropriate
solvents and, in most cases, in aqueous solvents, dissociate
into a macroion and low-mass ions (counterions)
\cite{Hara93,Schmitz93}. In spite of their ubiquitous presence
in technological and mostly biological processes, the complex
phenomenology they present is still far to be completely
understood and still eludes a complete description from a
theoretical point of view. The dynamical properties of these
systems, as manifested in electrophoretic mobility, diffusion,
viscosimetry and transport processes under the influence of
external driving forces, are deeply different from those of
neutral polymers. Their peculiar behavior is mainly due to the
coupling between electrostatic interactions among various
charged units along the polyion chain, that gives rise to a very
complex phenomenology whose understanding is still rather poor.

As far as the transport properties are concerned, the peculiar
feature which governs most of the polyion behavior is the
counterion condensation effect, for which the effective charge
of a polyion chain differs from the one based on a
stoichiometric ionization of its charged groups. Because of a
fine balance between electrostatic attraction to the polyion
backbone and the loss in the conformational entropy when
counterions are released in the bulk of the solution, a fraction
of counterions remains localized in the immediate vicinity of
the polyion chain (counterion condensation). As pointed out by
Jeon and Dobrynin \cite{Jeon07}, the situation is even more
complicated in case of flexible, or partially flexible polyions,
where the conformation of the chain, besides the concentration
regime, depends on the quality of the solvent. While for
$\theta$ or good solvents for the polyion backbone, counterion
condensation results in a decrease of the polyion size with the
increase of concentration, in poor solvent condition, where the
chain is modeled by a necklace of beads connected by strings of
monomers, condensation occurs in an avalanche-like fashion
\cite{Dobrynin2,Essafi, Chepelianskii}.

A complete approach to these systems within the theory of polymer
solution is rather difficult because of the intrinsical asymmetry of
a polyelectrolyte solution, where we can roughly identify a polymer
domain, where a highly charged flexible or partially flexible
polyion chain keeps a fraction of counterions, and a counterion
domain formed by small ions (from the dissociation of the ionizable
groups) and, in some cases, by cations and anions carrying one or
few charges (from the dissociation of added simple salt). Moreover,
while in the polymer domain the charges are partially correlated
along the polyion backbone, in the counterion domain charges can
freely or almost freely diffuse.

Most of the structural and dynamical properties of
polyelectrolytes, mainly in a biological context, are controlled
by the charge density along the polyion chain as well as by the
ionic strength of the surrounding aqueous phase. The knowledge
of the transport parameters associated with the ion distribution
inside the solution, both those bound to the immediate vicinity
of the polyion and the more free ones in the bulk solution, is
therefore particularly relevant in describing the behavior of
the polyion chain in solution.

It is understandable then how electrical conductivity, that
measures the contribution of every charged entity under the
influence of an external electric field, represents a valuable
probe to investigate the polyion behavior in different
environmental conditions.

We present here a set of low-frequency electrical conductivity (d.c.
conductivity) measurements of polyions of different chemical
structures, different flexibility and different molecular weights in
aqueous solutions (in absence of added salt), over an extended
concentration range, covering both the dilute and the semidilute
regime. The results are presented and discussed in the light of the
scaling theory of polyelectrolyte solutions.

We have chosen polyions with main chains of different flexibility,
such as carboxymethylcellulose sodium salt of two different
molecular weights, whose backbone is rather rigid, and such as
polycrylate, in salty and acidic form, and polystyrene sulfonic
acid, whose backbones are largely flexible, even if at a different
extent.

The paper is organized as follows. In Sec. II, we review the
basic equations dealing with the electrical conductivity of a
polyion solution and summarize the main predictions of the
scaling theories of a polyelectrolyte solution according to the
models proposed by Dobrynin et al. \cite{Dobrynin95} for
different concentration regimes. In Sec. III, we hint to the
chemical characterization of the polyions investigated and in
Sec. IV we present some experimental results for the
low-frequency electrical conductivity of the four different
class of polyions in an extended concentration range. In
particular, we present the behavior of the effective charge of
the polyion chain or, equivalently, the fraction $f$ of free
counterions, in the different polyion concentration regimes.
Finally, Sec. IV concludes with a brief discussion.

\section{\label{sec:level1}Theoretical background}

An ensemble of charged particles at a numerical concentration $n_i$,
each of them bearing an electric charge $z_ie$ and moving under the
influence of an external electric field with a mobility $u_i$, is
characterized by an overall electrical conductivity $\sigma$ given
by
\begin{equation}\label{eq.1}
    \sigma=\sum_i(|z_i|e)n_iu_i
\end{equation}
In the case of a polyion solution, in the absence of added salt,
because of the partial ionization of the polyion charged groups
(and the consequent counterion condensation effect), polyion
chains (at a concentration $n_p$, with a mobility $u_p$) and
released counterions (at a concentration $n_1$ and charge
$z_1e$) will contribute to eq. \ref{eq.1}, according to the
relationship
\begin{equation}\label{eq.2}
    \sigma=Z_{pol}en_pu_p+z_1en_1u_1
\end{equation}
where $Z_{pol}$ and $z_1$ are the valencies of the polyion chain
and a counterion, respectively.

Consider a polyion with degree of polymerization $N$, size of
the monomers $b$ and fraction of charged monomers $f$. The
Manning model \cite{Manning69,Manning78} predicts a
re-normalization of the effective charge of the polyion chain to
a constant value above a condensation threshold. According to
this theory, for a charged infinite linear polyion in very
dilute condition, a complete ionizable groups dissociation
occurs only when electrostatic repulsion is weaker than the
thermal energy $K_BT$; in other words, when the average distance
$b$ between charges along the chain is larger than the Bjerrum
length $l_B=e^2/\epsilon K_BT$ (with $e$ the elementary charge
and $\epsilon$ the permittivity of the aqueous phase). When $b$
is smaller than $l_B$, electrostatic interactions increase and a
fraction of counterions remains bound in the neighboring of the
polyion chain (counterion condensation). In the Manning theory,
counterions condense until $l_B/b_{eff}=1$, where $b_{eff}$ is
the effective distance between dissociated groups. Consequently,
the fraction ($1-f$) of counterions bound to the polyion chain,
in order to reduce its effective charge to the critical value,
will given by $1-b/l_B$.

Because of the counterion condensation effect, the charge of each
chain is $Q_p=Z_{pol}e=fNz_pe$ and each chain will release into the
aqueous phase $N f \nu_1$ counterions (each of them of charge
$Q_1=z_1e$). Electroneutrality of the whole solution requires
$z_p=\nu_1 z_1$ and eq. \ref{eq.2} reads
\begin{equation}\label{eq.3}
    \sigma=N f e n_p(z_pu_p+z_1\nu_1u_1)
\end{equation}
The mobility $u_1$ of counterions present in the solution is
given by Manning \cite{Manning69, Manning3}, according to
\begin{equation}\label{eq.4}
    u_1=u_1^0\frac{D_1}{D_1^0}-u_p(1-\frac{D_1}{D_1^0})
\end{equation}
where $u_1^0$ is the limiting mobility in the pure aqueous phase
and $D_1/D_1^0$ is the ratio of the self-diffusion coefficient
$D_1$ of counterions in the polyion solution and its value
$D_1^0$ in the pure solvent. Substitution of eq. \ref{eq.4} into
eq. \ref{eq.3} results
\begin{equation}\label{eq.5}
    \sigma=N f e n_p z_1\nu_1\frac{D_1}{D_1^0}(u_1^0+u_p)
\end{equation}
This equation, written in the usual notations, by introducing
the polyion concentration $C_p$ in mol/l and the equivalent
conductance of the polyion chain $\lambda_p=u_pF$, with $F$ the
Faraday constant, results
\begin{equation}\label{eq.6}
    \sigma=NfC_p\nu_1z_1\frac{D_1}{D_1^0}(\lambda_1^0+\lambda_p)
\end{equation}

The equivalent conductance $\lambda_p$, taking into account the
asymmetry field effect \cite{Bordi02b,Kirkwood}, can be written
as
\begin{equation}\label{eq.7}
    \lambda_p=\frac{FQ_p\frac{D_1}{D_1^0}}{f_E+\frac{Q_p}{u_1^0}(1-\frac{D_1}{D_1^0})}
\end{equation}
where $f_E$ is the usual electrophoretic coefficient (without
the asymmetry field correction). This expression holds both when
counterion condensation occurs and when does not, the two case
being differentiated by the value of the polyion charge $Q_p$.

Finally, the electrophoretic coefficient $f_E$ for a chain of
$N_0$ simple spherical structural units of radius $R_0$,
following the general expression given by Kirkwood \cite{Kirkwood}
can be written as
\begin{equation}\label{eq.8}
    f_E=\frac{N_0\zeta_{R_0}}{1+\frac{\zeta_{R_0}}{6\pi\eta N_0}\sum_i\sum_{i\neq j}
    \langle r_{ij}^{-1}\rangle}=\frac{N_0\zeta_{R_0}}{1+\frac{\zeta_{R_0}}{3\pi\eta R_0}
    |\ln(k_DR_0)|}
\end{equation}
where $\zeta_{R_0}=3\pi \eta R_0$ is the friction coefficient,
$\eta$ the viscosity of the aqueous phase and $r_{ij}$ the
distance between different structural units and $k_D=\sqrt{4\pi
l_b f c z_1^2}$ the inverse of the Debye screening length. The
sum in eq. \ref{eq.8} has been evaluated by introducing an
appropriate cut-off function $\exp(-k_D r_{ij})$ and a full
extended polyion chain with $r_{ij}=|i-j|R_0$ \cite{Bordi02}.

The electrophoretic coefficient $f_E$ depends on the particular
conformation assumed by the polyion chain in the considered
concentration regime, through the parameters $N_0$ and $R_0$,
and this quantity must be evaluated differently in the light of
the different flexibility of the polyion chain and of the
scaling model of polyion solutions.

\subsection{Flexible polyions: the scaling approach}

If we confine ourselves to flexible polyions in good solvent
conditions, in dilute solution, the polyion chain is modeled as
an extended rod-like configuration of $N_D=N/g_e$ electrostatic
blob of size $D$, each blob containing $g_e$ monomers and
bearing an electric charge $q_D=z_pefg_e$. In the semidilute
regime, the polyion chain is modeled as a random walk of
$N_{\xi_0}=N/g$ correlation blobs of size $\xi_0$, each
correlation blob containing $g$ monomers and bearing an electric
charge $q_{\xi_0}=z_pefg$. Following the procedure adopted by
Manning \cite{Manning81}, in dilute regime, the elementary unit
is the electrostatic blob and eq. \ref{eq.8} becomes
\begin{equation}\label{eq.9}
    f_E=\frac{N_D\zeta_D}{1+\frac{\zeta_D}{3\pi\eta D}|\ln(1/N_D)|}\simeq
    \frac{3\pi\eta N_D D}{\ln(N_D)}
\end{equation}
where the friction coefficient $\zeta_D$ is simply given by
\begin{equation}\label{eq.10}
    \zeta_D=3\pi\eta D
\end{equation}
In the semidilute regime, where the single unit is the correlation
blob of size $\xi_0$, the electrophoretic coefficient $f_E$ for a
random walk of $N_{\xi_0}$ correlation blobs is given by
\begin{equation}\label{eq.11}
    f_E=\frac{N_{\xi_0}\zeta_{\xi_0}}{1+\frac{8\sqrt{N_{\xi_0}}\zeta_{\xi_0}}{3\sqrt{6\pi^3 \eta \xi_0}}}
\end{equation}
where the friction coefficient $\zeta_{\xi_0}$ of the single
rod-like structure of size $\xi_0$ can be derived from eq.
\ref{eq.9} with the substitution of $N_DD$ with $\xi_0$ and $N_D$
with $N_D/N_{\xi_0}=g/g_e$,
\begin{equation}\label{eq.12}
    \zeta_{\xi_0}=\frac{3\pi\eta \xi_0}{\ln(g/ge)}
\end{equation}

\subsubsection{The scaling laws in the scaling polyion model}
In the dilute regime, the polyion equivalent conductance
$\lambda_p$ (eq. \ref{eq.7}) depends through the expression for
the electrophoretic coefficient $f_E$ on the length of the full
extended chain $N_DD$ and on the number $N_D$ of blobs in each
chain. According to Dobrynin et al. \cite{Dobrynin95}, these
quantities scale with the fraction $f$ of free counterions as
\begin{equation}\label{eq.12}
    N_DD\sim Nb(l_B/b)^{2/7}f^{4/7}
\end{equation}
\begin{equation}\label{eq.13}
    N_D\sim N(l_B/b)^{5/7}f^{10/7}
\end{equation}
In the semidilute regime, the equivalent conductance $\lambda_p$
(eq. \ref{eq.7}), again through the expression for the
electrophoretic coefficient $f_E$, depends on the contour length
$N_{\xi_0}\xi_0$ of the random walk chain of correlation blobs,
on the number $N_{\xi_0}$ of correlation blobs within each
polymer chain and on the ratio $g/g_e$ of the monomers inside a
correlation blob to those inside an electrostatic blob. Again,
according to Dobrynin et al. \cite{Dobrynin95}, these quantities
scale as
\begin{equation}\label{eq.14}
    N_{\xi_0}\xi_0\sim Nb(l_b/b)^{2/7}f^{4/7}
\end{equation}
\begin{equation}\label{eq.15}
    N_{\xi_0}\sim N b^{3/2}c^{1/2}(l_B/b)^{3/7}f^{6/7}
\end{equation}
\begin{equation}\label{eq.16}
    g/g_e\sim b^{-3/2}c^{-1/2}(l_B/b)^{2/7}f^{4/7}
\end{equation}

\subsection{Rigid polyions}
In the case of less flexible polyions, where the chain cannot be
modeled as a sequence of electrostatic or correlation blobs, the
electrophoretic coefficient $f_E$ (eq. \ref{eq.8}), and hence the
electrical conductivity, depends on the number $N_0$ of structural
units along the chain and on their size $R_0$. In the dilute regime,
these two quantities can be identified with the degree of
polymerization $N$ and with the monomer size $b$, respectively. In
the semidilute regime, the Khun formalism \cite{Yamakawa} can be applied
and the conformation of the polymer chain can be described in terms
of a statistical chain of $N_{l_k}$ segments, each of them of length
$l_k$. These parameters can be estimated through the persistence
length $L_p$ and the end-to-end distance $R_{ee}$ by means of the
relationships
\begin{equation}\label{}
    l_k=2L_p
\end{equation}
and
\begin{equation}\label{}
    R_{ee}=\sqrt{\langle(R(L)-R(0))^2\rangle}=(N_{l_k})^{1/2}l_k
\end{equation}
A summary of the different quantities involved in the
conductivity behavior of the polyion solutions, depending on the
flexibility of the chain and on the concentration regime is
given in Tab. \ref{Tab.1} These quantities have been employed in
the electrical conductivity analysis.
\begin{table}
  \centering
  \begin{tabular}{c c c c c}
\\
\hline
Structure & \hspace{0.2cm} Dilute regime &  \hspace{0.2cm} Semidilute regime\\
\hline \hline Flexible & \hspace{0.2cm} $N_D$, $D$ (eqs.
\ref{eq.7}, \ref{eq.9})
&  \hspace{0.2cm}  $N_{\xi_0}, \xi_0$ (eqs. \ref{eq.7}, \ref{eq.11}) \\
Rigid & \hspace{0.2cm} $N_b, b$ (eqs. \ref{eq.7}, \ref{eq.8})&
\hspace{0.2cm} $N_{l_k}, l_k$ (eqs. \ref{eq.7}, \ref{eq.8}) \\
\hline \hline
\end{tabular}
  \caption{Number and type of electrostatic units
  employed in the conductivity analysis. $b$ is the mean distance between
  charges, $l_k$ is the Khun length, $\xi_0$ is the correlation length
  of the polyions.}\label{Tab.1}
  \end{table}

\subsection{Different concentration regimes}

\begin{figure}[htbp]
\begin{center}
  \includegraphics[width=8cm]{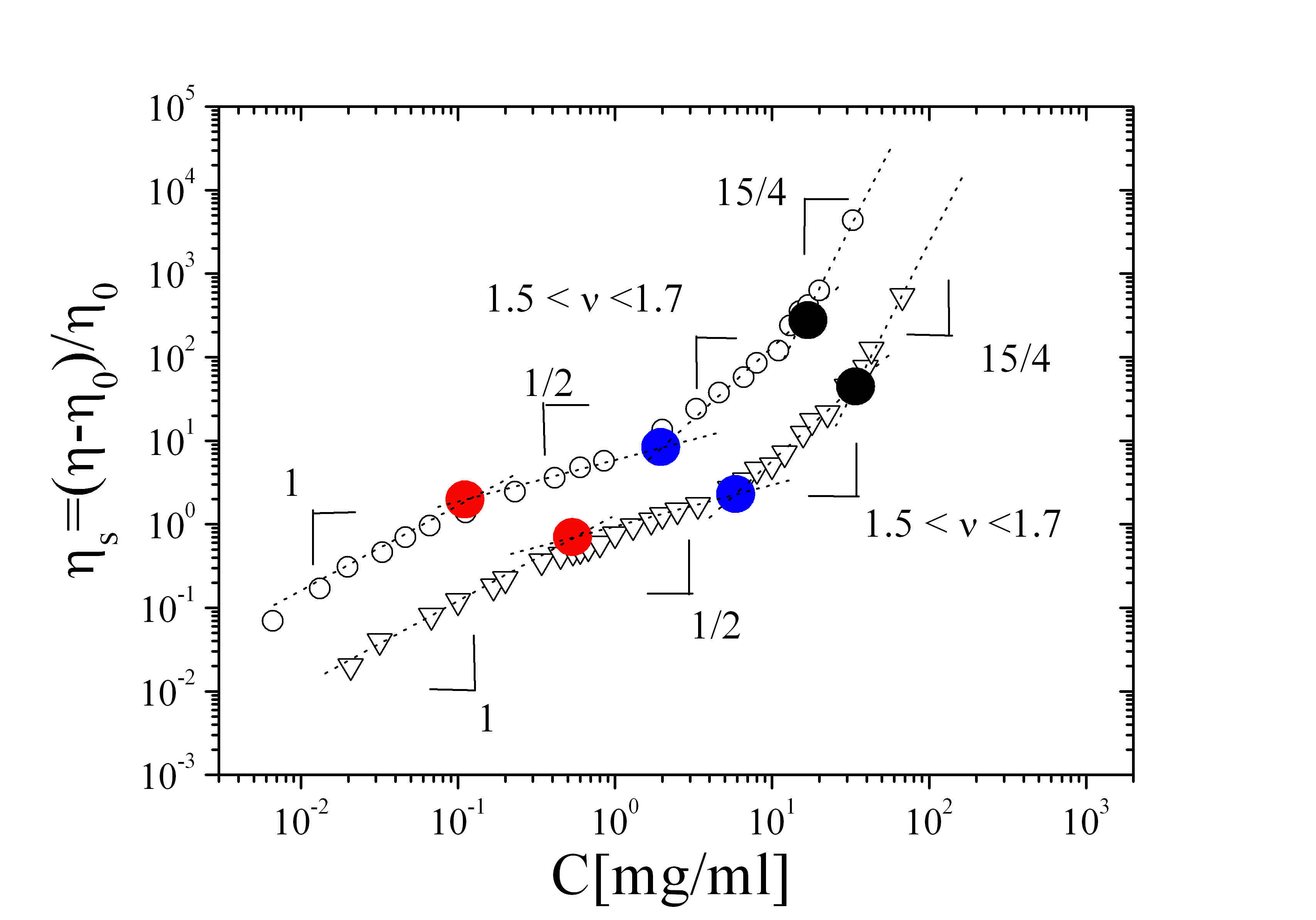}
  \caption{Reduced viscosity $\eta_s$ as a function of polyion concentration
  for Na-CMC-90 kD ($\bigtriangledown$) and Na-CMC-700 kD ($\bigcirc$). The dotted lines represent the expected
  power laws behaviors according to the different concentration regimes
  as shown in Tab. \ref{Tab.2}. The thick dots
  mark the boundary concentrations between dilute and semidilute unentangled ($c^*$),
  semidilute unentangled and semidilute entangled ($c_e$), semidilute entangled and
  concentrated ($c_D$) regimes, respectively.}\label{CMC_v}
\end{center}
\end{figure}

\begin{figure}[htbp]
\begin{center}
  \includegraphics[width=7cm]{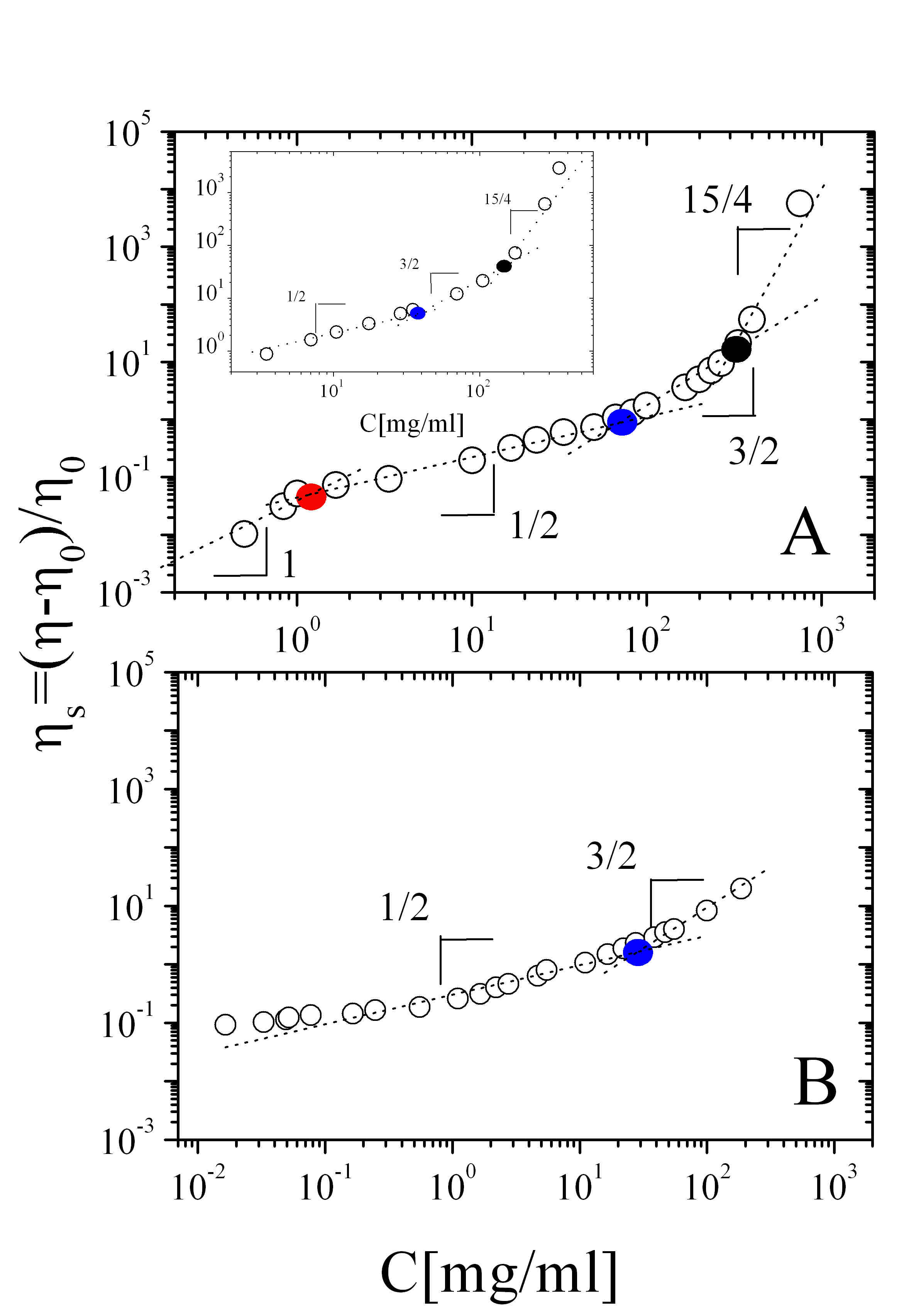}\\
  \caption{Reduced viscosity $\eta_s$ as a function of polyion
  concentration
  for NaPSS-Mal 20 kD (panel A) and H-PSS 20 kD (panel B). The dotted lines
  represent the expected power law behaviors according to the different
  concentration regimes as shown in Tab. \ref{Tab.2}. The thick dots
  mark the boundary concentrations between dilute and semidilute unentangled ($c^*$),
  semidilute unentangled and semidilute entangled ($c_e$), semidilute entangled and
  concentrated ($c_D$) regimes, respectively. The inset in panel A shows the
  reduced viscosity of Na-PAA (60 kD molecular weight) aqueous
  solution.
  }\label{PSS_v}
\end{center}
\end{figure}

The behavior of polyelectrolyte aqueous solutions is
characterized by different concentration regimes governed by
four different zero-shear viscosity scaling laws \cite{Dobrynin,
Muthu2}. We can identify three typical transition concentration
thresholds, the concentration $c^*$, above which chains begin to
overlap, the concentration $c_e$, at which entanglement effects
begin and, finally, the concentration $c_D$, at which single
electrostatic units (blobs or single monomers) start to overlap.
For $c<c^*$ (dilute regime), each chain does not interact with
others and an extended chain conformation is expected and the
Zimm dynamics reasonably describes diffusive processes of
isolated chains. For $c^*<c<c_e$ (non entangled semidilute
regime), excluded volume and hydrodynamics interactions are
screened and entanglement effects are weak. In this
concentration regime, the Rouse dynamics applies and typical
Fuoss law for the reduced viscosity is expected. For $c>c_e$
(entangled semidilute regime), entanglement effects dominates.
Significant overlap of chains occurs and topologically
constrained motion results in a reptation-tube diameter larger
than the correlation length \cite{Dobrynin}. Finally, for
$c>c_D$, a concentrated regime holds. These different
concentration regimes are characterized by different viscosity
behaviors. The expected power laws for zero-shear reduced
viscosity is given by $\eta_s=(\eta-\eta_0)/\eta_0 \sim
c_p^\nu$, where $\eta$ is the viscosity of polyion solution,
$\eta_0$ is the viscosity of the solvent and $c_p$ is the
polyion concentration. The expected exponents $\nu$ of the
polyion concentration dependence are listed in Tab. \ref{Tab.2}.
\begin{table}
  \centering
  \begin{tabular}{c c}
reduced viscosity $\eta_s$ & scaling exponent $\nu$\\
\hline \hline
$c<c^*$& $1$\\
$c^*<c<c_e$& $1/2$\\
$c_e<c<c_D$& $1.5 \div 1.7$\\
$c>c_D$& $15/4$\\
\hline \hline
\end{tabular}
  \caption{Expected scaling exponents for zero-shear reduced
  viscosity \cite{Dobrynin, Muthu2} of aqueous polyion solutions in the four typical concentration
  regimes.}\label{Tab.2}
\end{table}
The reduced viscosity $\eta_s=(\eta-\eta_0)/\eta_0 $ of the
different aqueous polyion solutions investigated has been
measured in order to evidentiate the different concentration
regimes, with the further aim of determining, as accurately as
possible, the boundary concentrations between the different
regimes. Figs. \ref{CMC_v} and \ref{PSS_v} show the reduced
viscosity $\eta_s$ as a function of the polyion concentration.
The different scaling regimes, characterized by different
scaling exponents, are well evidenced. For all the systems, the
concentrations $c^*$, $c_e$ and $c_D$ are easily identified.
\section{Experimental}

\subsection{Materials}

Polyacrylate in salty form [Na-PAA] and in acidic form [H-PAA],
were purchased from Sigma Chem. Co. as  25 wt\% solutions in
water. Sodium salt of carboxymethylcellulose [Na-CMC] of two
different molecular weights (90 and 700 kD), Polystyrene
sulfonic acid [H-PSS] (18 wt\% in water solution) and
Poly(4-styrenesulfonic acid-co-maleic acid) sodium salt
[Na-PSS-MA] were purchased from Sigma-Aldrich. Co. The samples
were used as received, without any further purification.

The polymers differ by flexibility, molecular weight, linear
charge density. The sodium salt of poly(acrylic acid) is a water
soluble polyelectrolyte with a relatively simple chemical repeat
unit such that can be considered as a model flexible polyion.
Na-CMC molecules assume a rather extended, rod-like,
conformation at low concentrations but, at higher
concentrations, the molecules overlap and coil up until they
behave as a thermoreversible gel. PSS is probably the most
widely studied synthetic polyelectrolyte, with a rather flexible
chain at room temperature. The main structural and chemical
characteristics of the polymers investigated are listed in Tab.
\ref{Tab.3}.

All the solutions were prepared at the desired polymer concentration
(in the range from 10$^{-2}$ mg/ml to 100 mg/ml, in order to
completely cover the dilute and semidilute concentration regime)
with deionized Q-quality water (Millipore) whose d.c. electrical
conductivity was less than 1$\div$2 $\mu$mho/cm at room temperature.
\begin{figure}
   \begin{center}
      \includegraphics*[width=3.0in]{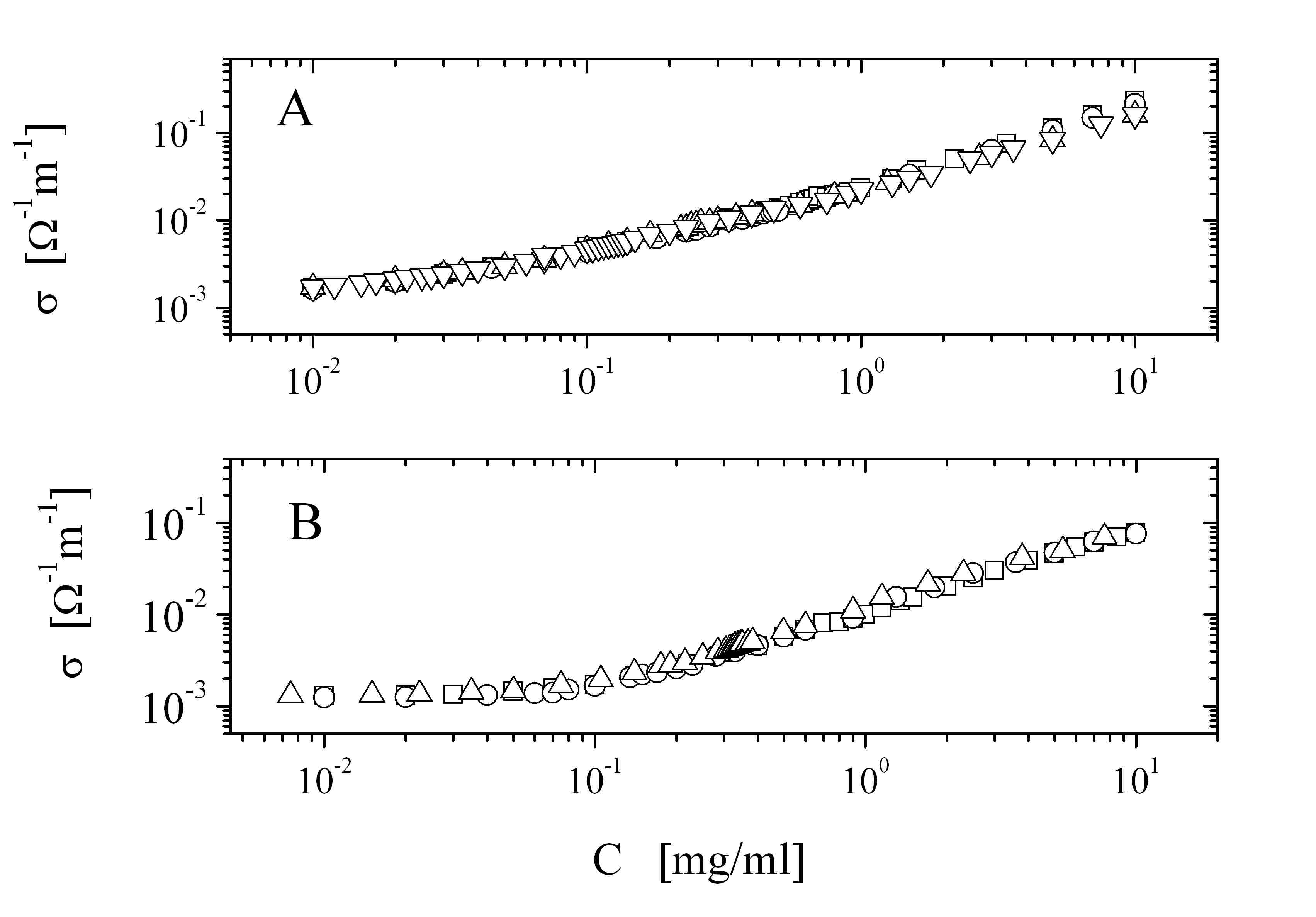}
      \caption{The low-frequency electrical conductivity $\sigma$ of
      polyacrylate 60 kD molecular weight
      both in salty form [Na-PAA] (panel A) and in acidic form
      [H-PAA] (panel B) as a function of the polyion concentration,
      at the temperature of 25.0 $^{\circ}$C. Different symbols refer to
      three independent set of measurements on independent sample preparations.}
      \label{fig:Fig_1}
   \end{center}
\end{figure}
\begin{figure}
   \begin{center}
      \includegraphics*[width=3.0in]{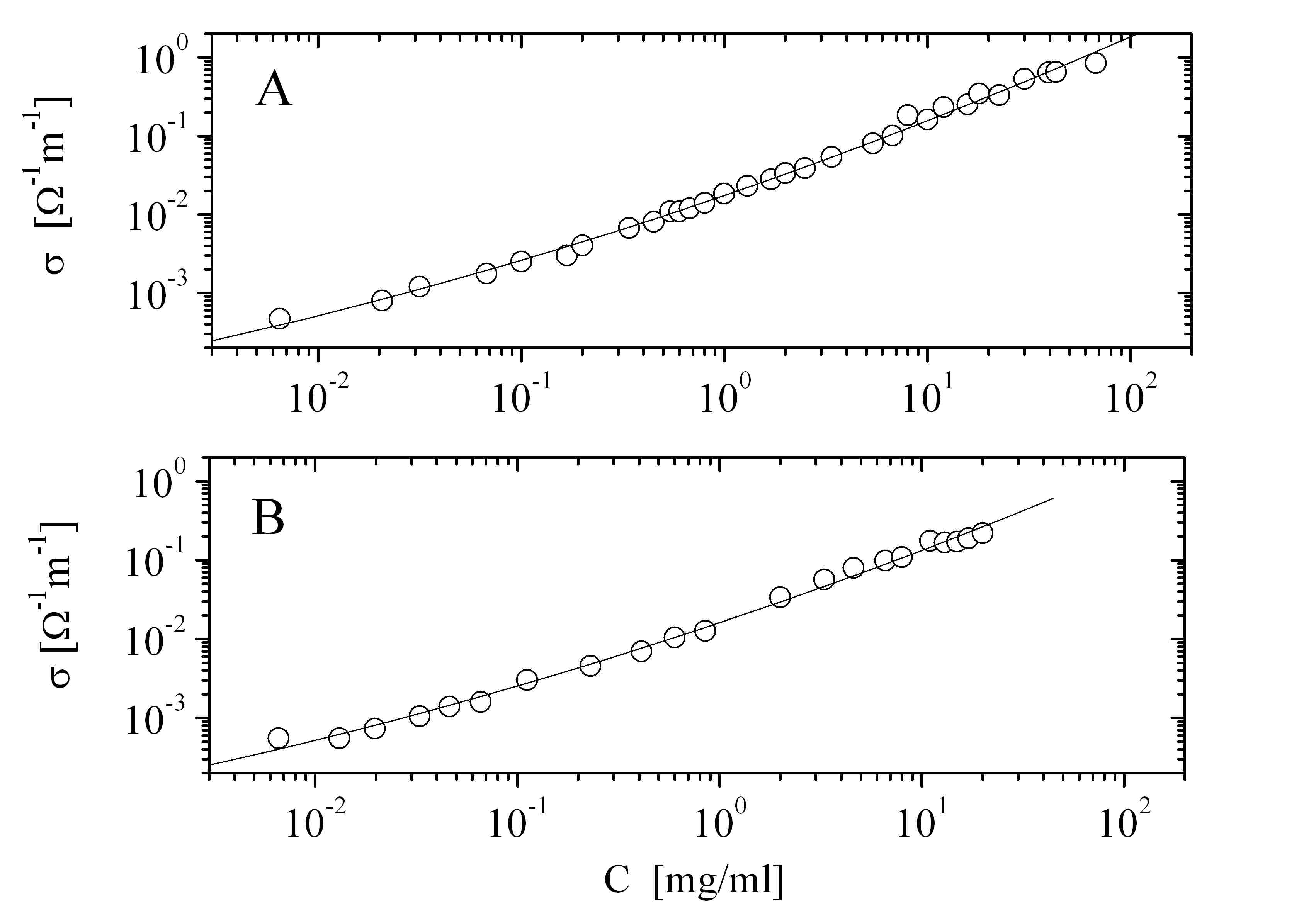}
      \caption{The low-frequency electrical conductivity $\sigma$ of
      Na-CMC as a function of the polyion concentration,
      at the temperature of 25.0 $^{\circ}$C, at two different molecular
      weight: (panel A): 90 kD; (panel B): 700 kD }
      \label{fig:Fig_2}
   \end{center}
\end{figure}
\begin{figure}
   \begin{center}
      \includegraphics*[width=3.0in]{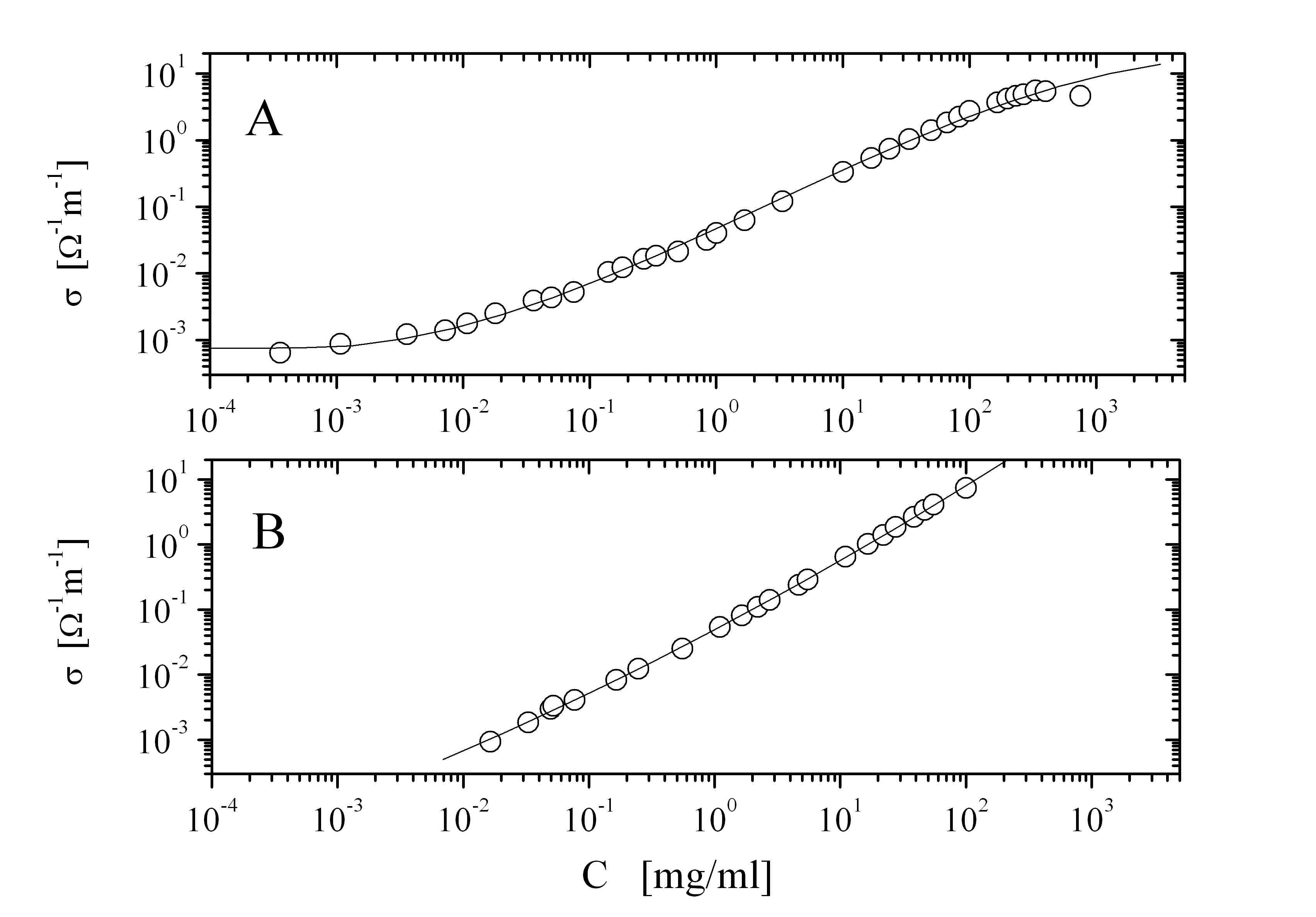}
      \caption{(panel A): The low-frequency electrical conductivity $\sigma$ of
      Na-PSS-MAL 20 kD molecular weight as a function of the polyion concentration,
      at the temperature of 25.0 $^{\circ}$C.
      (panel B): the analogous quantity for H-PSS 75 kD molecular weight.}
      \label{fig:Fig_3}
   \end{center}
\end{figure}

\subsection{Electrical conductivity measurements}
The electrical conductivity measurements were carried out by
means of two Impedance Analyzers Hewlett-Packard mod. 4192A over
the frequency range from 40 Hz to 110 MHz and mod. 4191A in the
frequency range from 1 MHz to 2 GHz. All measurements have been
carried out at the temperature of 25.0 $^\circ$C within 0.1
$^\circ$C. The conductivity cell consists in a short section of
a cylindrical coaxial cable connected to the Meter by means of a
precision APC7 connector, whose constants have been determined
by a calibration procedure with standard liquids of known
conductivity and dielectric constant, according to a procedure
reported elsewhere \cite{Bordi97,Bordi01}. Here, we will deal
exclusively with the low-frequency limit of the measured
electrical conductivity, obtained by averaging the conductivity
values at frequencies below 10 kHz, where only the d.c.
contributions are present, being completely negligible the
contributions due to the dielectric loss (d.c. conductivity).
The behavior of the electrical conductivity as a function of the
frequency will be presented and discussed in a forthcoming paper
\cite{Truzzolillo08}.

\subsection{Polyelectrolyte behavior}

All the polyion chains have at least one ionizable group per
monomer with a mean distance $b$ between charges (Tab.
\ref{Tab.3}). In aqueous solution, the charge density parameter
$l_b/b$ is larger than the critical value ($l_b/b=1$) and
according to the Manning-Oosawa theory \cite{Manning1},
counterion condensation will occur. However, it is worth noting
that Manning condensation doesn't apply in extremely dilute
polyelectrolyte solutions \cite{Wandrey1, Wandrey2, Muthu1,
Lin}, because of the huge entropy gain occurring when
single-chain system passes from a counterion-condensed state to
a counterion-uncondensed state.
\begin{table}[h]
  \centering
  \begin{tabular}{c c c c c c}
\\
\hline
     & $M_w[kD]$ & $N$ & $b[{\AA}]$ & $L_p$ [\AA] &   Structure\\
\hline \hline
CMC (I) & $90$ & $400$ & $6.87$ & $\simeq 160$ &   rigid\\
CMC (II) & $700$ & $3200$ & $6.87$ & $\simeq 160$ &   rigid\\
Na-PAA & $60$ & $638$ & $2.52$ & $\simeq 7$ &   flexible\\
H-PAA & $60$ & $638$ & $2.52$ & $\simeq 7$ &   flexible\\
H-PSS & $75$ & $407$ & $2.50$ & $\simeq 10$ &   flexible\\
Na-PSS-Mal & $20$ & $103$ & $3.30$ & $\simeq 10$ &   flexible\\
\hline \hline
\end{tabular}
  \caption{Molecular weight $M_w$, degree of polymerization $N$,
  persistence length $L_p$ (taken from ref. \cite{Hoogendam, Schiessel, Manning06}),
  monomer size $b$ and flexibility
  properties of the polyions investigated.}\label{Tab.3}
\end{table}

\section{Results and discussion}

The low-frequency electrical conductivity $\sigma$ of the
different polyion aqueous solutions investigated are shown in
Figs. \ref{fig:Fig_1} to \ref{fig:Fig_3}. Measurements extend
over a wide range of polyion concentration, covering the dilute
and semidilute regime, up to the beginning of concentrated
regime. As can be seen, when plotted in a double log-scale, the
conductivity $\sigma$ shows a marked power-law behavior, with
marked deviations both at low and high concentrations. These
deviations are more relevant in the case of the copolymer
Na-PSS-Mal and of the polymer H-PAA. The electrical conductivity
data have been analyzed on the basis of the above stated
theoretical framework, with the appropriate expressions for
polyion equivalent conductivity, according to the concentration
regime and the flexibility of the polyion chain.

For flexible polyelectrolytes, the elementary units are
electrostatic blobs and correlation blobs at low (dilute regime)
and high (semidilute regime) concentration, respectively. The
appropriate expressions for the polyion conductivity in dilute
and semidilute regimes are given by eqs. \ref{eq.6} and
\ref{eq.7}, where the electrophoretic coefficients are given by
eq. \ref{eq.9} for $c<c^*$ and by eq. \ref{eq.11} for $c>c^*$.
For rigid polyelectrolytes, we cannot consider the monomer
chains as linear sequence of blobs of size $D$ or a sequence of
correlation blobs of size $\xi_0$ and we will consider, as
simple electrostatic units, the segments of length $b$ and the
Khun length $l_k$ in dilute and semidilute regimes,
respectively. In other words, for semidilute flexible polymer
solutions, we will assume the chain composed by a sequence of
Khun segments, in which the Kuhn length plays the role of the
correlation length $\xi_0$ in the case of more flexible polyions
within the semidilute regime.

Eqs. \ref{eq.6} and \ref{eq.7} for the electrical conductivity in
dilute and semidilute regime depend on a single free parameter, the
fraction $f$ of free counterions, the other parameters being known
through the structural properties of the polyion backbone or assumed
to be known within the theoretical framework employed. By means of a
non-linear least-squares fitting procedure of the above stated
models to the experimental results, the values of $f$, over the
whole concentration range investigated, can be obtained.

\begin{figure}[htbp]
\begin{center}
  \includegraphics[width=7cm]{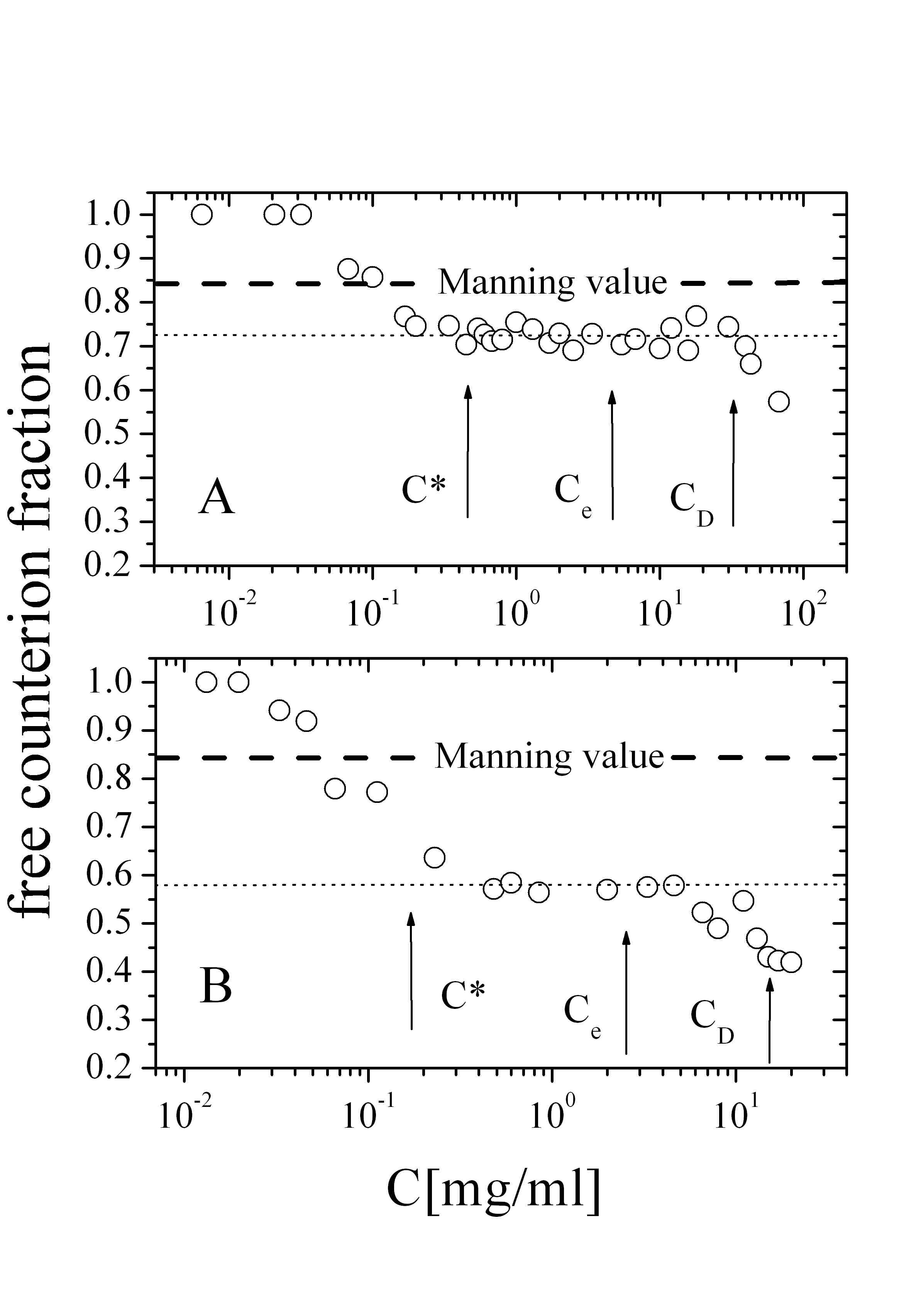}
  \caption{The fraction $f$ of free counterions in Na-CMC
  aqueous solutions as a function of the polymer concentration. Panel
  A: Na-CMC 90 kD; Panel B: Na-CMC 700 kD. The arrows mark transition
  concentrations as deduced from the viscosity measurements. Dashed
  line indicates $f$ value predicted by Oosawa-Manning theory while
  dotted line is the mean value $\langle f \rangle$ in the \emph{Manning regime}. }\label{CMC_f}
\end{center}
\end{figure}

\begin{figure}[htbp]
\begin{center}
  \includegraphics[width=8cm]{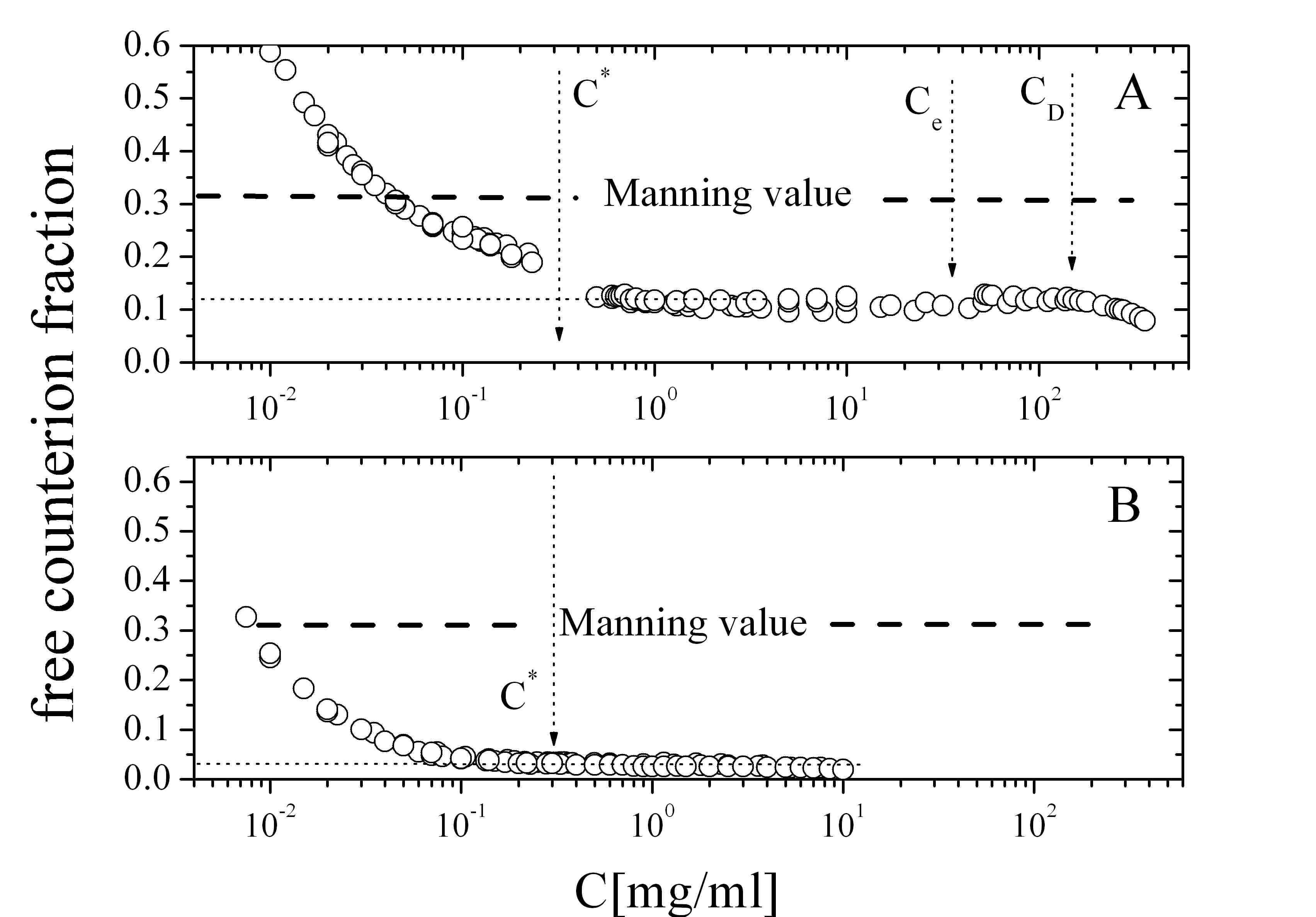}
  \caption{The fraction $f$ of free counterions
  in Na-PAA (Panel A) 60 kD and H-PAA 60 kD (Panel B) aqueous
  solutions as a function of the polymer concentration. The
  arrows mark transition concentrations as deduced from the
  viscosity measurements. Dashed line indicates $f$ value
  predicted by Oosawa-Manning theory while dotted line is
  the mean value $\langle f \rangle$ in the \emph{Manning regime}.
  }\label{PAA_f}
\end{center}
\end{figure}

\begin{figure}[htbp]
\begin{center}
  \includegraphics[width=8cm]{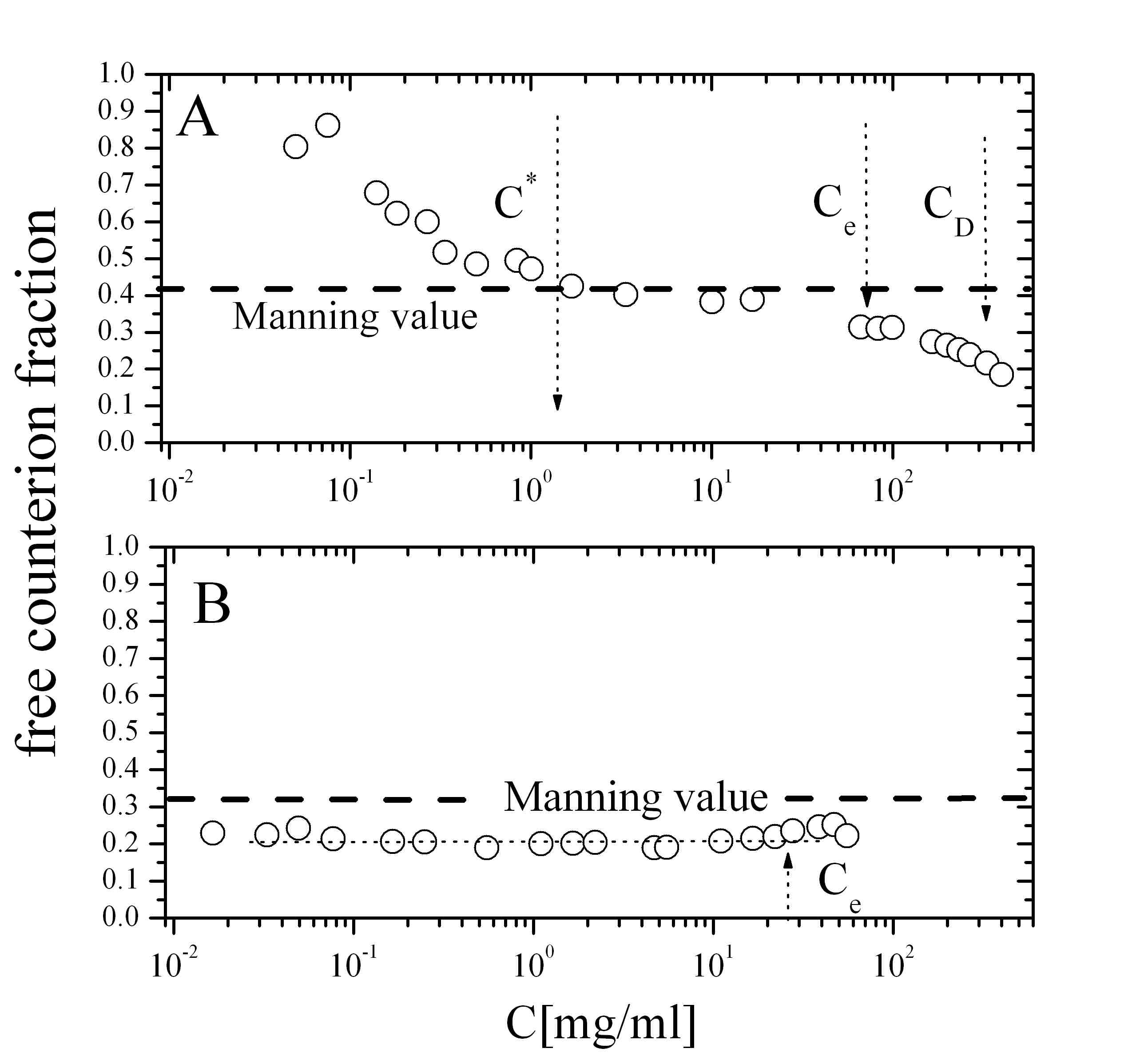}
  \caption{The fraction $f$ of free counterions in
  Na-PSS-Mal (Panel A) 20 kD and H-PSS 20 kD (Panel B) aqueous
solutions as a function of the polymer concentration. The arrows
mark transition concentrations as deduced from the viscosity
measurements. Dashed line indicates $f$ value predicted by
Oosawa-Manning theory while dotted line is the mean value $\langle f
\rangle$ in the the \emph{Manning regime}}\label{PSS_f}
\end{center}
\end{figure}

Figs. \ref{CMC_f} to \ref{PSS_f} show this parameter as a function
of polymer concentration $c$ for the different polymers
investigated. At a first look, for all the systems, there
is a pronounced decrease of $f$ as polymer concentration increases,
this decrease extending approximately over the whole dilute regime.
This finding is in agreement with previous theoretical and other
experimental results \cite{Muthu1, Liu, Lin, Cheng, Wandrey1, Wandrey2}.
For all the polyelectrolytes investigated and for semidilute
(non entangled) regime, $c^*<c<c_e$, a more or less extended
plateau is reached and we will name this phase as the
\emph{Manning regime} \cite{Wandrey1, Wandrey2}. This dependence
is of course unexpected within the classical Manning model, that
for a linear infinite rigid polyion, in the limit of infinite
dilution, predicts a fraction of free counterions fixed to the
value $f_{th}$ defined by the ratio between the nominal charge
spacing $b$ and Bjerrum length $l_b$ \cite{Manning1} according
to
\begin{equation}\label{mann1}
f_{th}=0.886|z_1|^{-1}\xi^{-1} \mathbf{\hspace{0.5cm}}
\xi=\frac{l_b}{b}
\end{equation}
where $z_1$ is the valence of counterions and $\xi$ is the
charge density parameter. While a constant value is observed,
its level is significantly lower than the one predicted by
Manning (eq. \ref{mann1}).

As far as the CMC polyions are concerned, we observe that the
level at which the fraction $f$ of free counterions maintains
constant, within the semidilute regime, depends on the degree of
polymerization $N$. This effect has been previously observed in
different other works \cite{Wandrey1,Wandrey3} in aqueous
polystyrene sulfonate sodium salt [Na-PSS] and in various
poly(vinyl-benzyl-trialkyl ammonium) chloride aqueous solutions.
Moreover, a concentration dependence of the fraction $f$ of free
counterions has been measured in a variety of polyion solutions
such as, for example,
in sodium polyacrylate salt 225 kD molecular weight in the
presence of added NaCl \cite{Bordi03c}, again in sodium
polyacrylate salt solution at high concentration
\cite{Bordi02,Bordi02b} and in differently structured compounds
of poly(vinyl-benyl-trialkyl ammonium) chloride or in sodium
poly(styrene sulfonate) \cite{Wandrey1,Wandrey2}. With the
increase of the degree of polymerization, the partial increase
of the chain flexibility and the partial overlap between
polyelectrolyte chains influence the amount of the net charge
$Nf$ of the single chains and determine the characteristic value
of the fraction $f$ of free counterions within the \emph{Manning
regime} that, as shown in Figs. \ref{CMC_f} to \ref{PSS_f}, has
always a lower value than the theoretical prediction (eq.
\ref{mann1}).
It is worth noting that the so called \textit{Manning regime} begins
at the concentration $c^*$, which defines the boundary between
dilute and semidilute regime. From viscosity measurements, these
concentrations fall at $c^*$=0.38 mg/ml and $c^*$=0.15 mg/ml for
CMC-90 kD and CMC-700 kD, respectively. These concentrations agree
reasonably well with the ones calculated from the relationship
$c^*=N/R_{ee}$, that yields values of about 0.3 mg/ml and 0.1 mg/ml,
respectively.

In the case of flexible polyions, i.e., Na-PAA, H-PAA,
Na-PSS-Mal and H-PSS polyions, the conductivity data have been
analyzed according to the scaling model of a polyion solution
and the resulting fraction $f$ of free counterions is shown in
Figs. \ref{PAA_f} and \ref{PSS_f}. As can be seen, also in these
cases, the same phenomenology appears, with a well-defined
constant interval for the $f$ values, corresponding in all the
cases investigated to the semidilute regime. For all the samples
investigated, at concentrations close to $c_D$, there is a
decrease of $f$, this behavior being as evident as in the case
of less flexible polyions.

For the sake of completeness, we remark that if we had extended
the Manning model to flexible polyions (either considering a
sequence of subunits of length $b$ or a sequence of subunits of
length $l_k$), an inconsistence would have been found, with
values of $f$ largely meaningless in extended concentration
ranges. This fact supports the need to use a different
theoretical framework (the scaling model of a polyion solution)
in the case of flexible polyions.

A final comment is in order. In the case of rigid polyions, we
have differently analyzed the conductivity data. If the model
adopted in the dilute regime had been extended to the semidilute
regime, considering the elementary unit as a segment of length
$b_{eff}$ and the number of segment given by $Nb/b_{ff}$, the
results, as far as the fraction $f$ of free counterions,
wouldn't have been sensibly different from the ones obtained
(Fig. \ref{CMC_f}) considering a change in the length scale of
basic electrostatic unit from the monomer length $b$ to the Khun
length $l_k$. This result furnishes further support to the
analysis of conductivity data employing different models in
consequence of different conformation of the polyion in
solution.


\section{Conclusions}

In this paper, taking advantage of the scaling picture of a
polyelectrolyte solution according to the theory developed by
Dobrynin, Colby and Rubinstein \cite{Dobrynin94,Dobrynin95}, we
present, for a series of differently flexible polyion solutions, how
the fraction of free counterions, due to the counterion
condensation, varies with the polyion concentration. These results
offer further support to the fact that the effective charge of a
polyion chain, after counterion condensation has occurred, depends
on the polymer concentration regime. In other words, the
distribution of free counterions in the bulk solution is influenced
by the presence of neighboring chains and, even more, by the
conformation they assume in the different concentration regime.

Furthermore, we have observed the presence of a \emph{Manning
regime}, where the fraction $f$ of free counterions remains
practically constant, when it coincides with the unentangled
semidilute regime. In this regime, the value of the free
counterion fraction is lower than the one predicted by classical
Manning theory at finite concentration \cite{Manning2}. On the
other hand, for all the polyions investigated, the dilute regime
is characterized by monotonic decrease of free counterion
fraction as polyion concentration is increased in accordance
with previous theoretical results \cite{Muthu1, Liu} that
strongly underly the importance of entropic effects at infinite
dilution.

The influence of the different flexibility chain effects have
been taken into account using the scaling blob model
\cite{Dobrynin} to analyze flexible chains (Na-PAA, H-PAA, H-PSS
and Na-PSS-Mal polyions) and a simple Manning scheme for rigid
polyions (CMC-90 kD and CMC-700 kD). At high concentration
regime, for $c>c_e$ and for all polyions investigated, $f$ shows
a further decrease that is more pronounced in the proximity of
the concentrated regime ($c>c_D$), where single electrostatic
units start to overlap and where single counterion is somehow
trapped between chains so that it can not contribute to the
specific conductivity of the solution. These counterions lead to
a sharp decrease of free counterion fraction.

The above analysis is for the time being confined to polyions in
good solvent condition, but it could be extended to polyions in
poor solvent conditions, where a strong re-structuring of the
polyion chain within the necklace model \cite{Dobrynin2} is
expected. These investigations may give useful information on
the conformation of a polyion in solution in different
concentration regimes and offer further insight on the mechanism
of the electrical conductivity of polyelectrolyte aqueous
solutions.


\bibliography{Pharma}

\end{document}